\documentstyle[12pt]{article}
\newlength{\extraspace}
\newlength{\extraspaces}
\newcommand{\be}{\begin{equation}
\addtolength{\abovedisplayskip}{\extraspaces}
\addtolength{\belowdisplayskip}{\extraspaces}
\addtolength{\abovedisplayshortskip}{\extraspace}
\addtolength{\belowdisplayshortskip}{\extraspace}}
\newcommand{\ee}{\end{equation}}
\newcommand{\ba}{\begin{eqnarray}
\addtolength{\abovedisplayskip}{\extraspaces}
\addtolength{\belowdisplayskip}{\extraspaces}
\addtolength{\abovedisplayshortskip}{\extraspace}
\addtolength{\belowdisplayshortskip}{\extraspace}}
\newcommand{\ea}{\end{eqnarray}}
\begin{document}
\thispagestyle{empty}
\setlength{\baselineskip}{6mm}
\begin{flushright}
SIT-LP-11/12 \\
December, 2011
\end{flushright}
\vspace{7mm}
\begin{center}
{\large \bf Nonlinear SUSY General Relativity Theory \\[2mm]
and Significances
}
\footnote{
Based on the talk given by K. Shima 
at the 7th International Conference {\it Quantum Theory and Symmetries}, 
07-13, August, 2011, Czech Technical University, Prague, Czech Republic
}
\\[20mm]
{\sc Kazunari Shima}
\footnote{
\tt e-mail: shima@sit.ac.jp} \ 
and \ 
{\sc Motomu Tsuda}
\footnote{
\tt e-mail: tsuda@sit.ac.jp} 
\\[5mm]
{\it Laboratory of Physics, 
Saitama Institute of Technology \\
Fukaya, Saitama 369-0293, Japan} \\[20mm]
\begin{abstract}
We show some consequences of the nonliear supersymmetric general relativity (NLSUSYGR) theory 
on particle physics, cosmology and their relations. They may give new insights into the SUSY breaking mechanism, dark energy, 
dark matter and the low enegy superpartner particles which are compatible with the recent LHC data. 
\\[5mm]
\noindent
%
%
%
\end{abstract}
\end{center}

\newpage

\section{Introduction}
Supersymmetry (SUSY) \cite{wb} and its spontaneous breakdown are profound notions essentially related to the space-time 
symmetry. Therefore it is natural to study them in the framework including the particle physics 
and the cosmology (gravitation) as well.  $SO(N)$ super-Poincar\'e (sP) symmetry may give a natural framework. 
We have found by the group theoretical arguments that among all $SO(N)$ sP groups only $SO(10)$ sP 
can accomodate in the low energy the standard model (SM) with just three generations 
of quarks and leptons in the single irreducible representation \cite{KS0},  
where we have adopted the decomposition $\underbar{10}_{SO(10)}=\underbar{5}_{SU(5)}+
\underbar{5}^{*}_{SU(5)}$ corresponding to $SO(10) \supset SU(5)$ and assigned to 
$\underline{5}_{SU(5)}$ the same quantum numbers as those of $\underline {5}$ of $SU(5)$ GUT \cite{KS}. 
Therefore it is an interesting problem to construct  $N=10$ SUSY theory in curved space-time. 
For this purpose we must overcome the so called no-go theorem of the $S$-matrix arguments in the local 
field theory for $N>8$ SUSY.  
We attempt to circumvent the no-go theorem by the degeneracy of the vacuum (local flat space-time). 
We consider new (unstable) four dimensional space-time whose tangent space is specified by 
the coordinates $SO(1,3)$ {\it and} $SL(2,C)$ suggested by nonlinear supersymmetry (NLSUSY). 
Extending the geometric arguments of the general relativity (GR) on ordinary Riemann space-time to new space-time 
we construct Einstein-Hilbert type action (NLSUSYGR theory) which is invariant under NLSUSY transformation. 
We discuss in this article some basic ideas and some consequences for 
the low energy particle physics and the cosmology of NLSUSY GR. 
We find that the self-contained phase transition of space-time  
plays crucial roles in our scenario. 
As the physical and the simplest case we consider $N=2$ case explicitly and show 
in the {\it true} vacuum of flat space-time that the lepton 
sector of SM with $U(1)$ gauge symmetry 
emerges as the composites of the fundamental Nambu-Goldstone(NG) fermion, 
i.e. the {\it true} vacuum of $N=2$ NLSUSY GR is achieved by the compositeness of particles of the SM.

\section{Nonlinear supersymmetric general relativity (NLSUSY GR)}
Nonlinear supersymmetric general relativity (NLSUSY GR) theory \cite{KSa} 
is based upon the general relativity (GR) principle
and the nonlinear (NL) representation \cite{VA} of SUSY \cite{WZ}.  
In NLSUSY GR,  four dimensional new space-time \cite{KSa}, as an ultimate shape of nature, is introduced, 
where tangent flat space-time has the NLSUSY structure, 
i.e. tangent space-time of four dimensional space-time manifold is specified by not only  the $SO(1,3)$ Minkowski coodinates $x_a$ 
but also  $SL(2,C)$ Grassman coordinates $\psi^i_\alpha$ ($i = 1, 2, \cdots, N$) for $N$-NLSUSY. 
The Grassmann coordinates in  new  space-time are regarded as
coset space coordinates of ${super GL(4,R) \over GL(4,R)}$,  
which can be recasted as the NG-fermions ({\it superon}) for NLSUSY  
associated with the spontaneous breakdown  of super-$GL(4,R)$ ($sGL(4,R)$) down to $GL(4,R)$. 
By extending the geometric arguments of Einstein GR in ordinary  Riemann space-time to new space-time, 
we obtain the fundamental action (NLSUSY GR theory) of the Einstein-Hilbert (EH) form \cite{KSa};  
\begin{equation}
L_{\rm NLSUSYGR}(w) = - {c^4 \over {16 \pi G}} \vert w \vert \{ \Omega(w) + \Lambda \}, 
\label{NLSUSYGR}
\end{equation}
where $G$ is the Newton gravitational constant, $\Lambda$ is a ({\it small}) cosmological constant 
indicating the NLSUSY structure in Riemann flat $e^a{}_\mu \rightarrow \delta^a{}_\mu$ space-time, 
$\Omega(w)$ is the unified Ricci scalar curvature of new space-time 
computed in terms of the unified vierbein $w^a{}_\mu$ (and the inverse $w_A^\mu{}$) defined by 
\begin{equation}
w^a{}_\mu = e^a{}_\mu + t^a{}_\mu(\psi), 
\ \ 
t^a{}_\mu(\psi) = {\kappa^2 \over 2i} 
(\bar\psi^i \gamma^a \partial_\mu \psi^i - \partial_\mu \bar\psi^i \gamma^a \psi^i), 
\label{unified-w}
\end{equation}
and $\vert w \vert = \det w^a{}_\mu$. 
In Eq.(\ref{unified-w}), $e^a{}_\mu$ is the ordinary vierbein of GR for the local $SO(3,1)$, 
$t^a{}_\mu(\psi)$ is the mimic vierbein  for the local $SL(2,C)$ $\psi^i(x)$ which is 
recasted as the stress-energy-momentum tensor 
of the NG fermion $\psi^i(x)$ created by the subsequent spontaneous breakdown of SUSY.       
$\kappa$ with the dimemsion (mass)$^{-2}$  is an arbitrary constant of NLSUSY representing 
the fundamental volume ($\kappa^2$) of four dimensional flat space-time which subsequently reinterpreted 
as the superon-vacuum coupling constant. 
Note that $e^a{}_\mu$ and $t^a{}_\mu(\psi)$ contribute equally to the curvature of space-time $\Omega(w)$, 
which realizes the Mach's principle in ultimate space-time. 

The $N$-NLSUSY GR action (\ref{NLSUSYGR}) possesses the following space-time and internal symmetries 
homomorphic to $SO(N)$ ($SO(10)$) sP symmetry \cite{ST1}, i.e. $L_{\rm NLSUSYGR}(w)$ is invariant under \\[2mm]
[{\rm new \ NLSUSY}] $\otimes$ [{\rm local \ GL(4,R)}] 
$\otimes$ [{\rm local \ Lorentz}] $\otimes$ [{\rm local \ spinor \ translation}] \\
\hspace*{2.5cm} 
$\otimes$ [{\rm global}\ SO(N)] $\otimes$ [{\rm local}\ U(1)$^N$] $\otimes$ [{\rm Chiral}]. \\[2mm]
Note that the no-go theorem is overcome (circumvented) in a sense that 
the nontivial $N$-extended SUSY gravity theory with $N > 8$ has been constructed in the NLSUSY invariant way, 
i.e. by the degeneracy of the vacuum (flat space-time). 

\section{Particle astrophysics of NLSUSY GR}
New {\it empty} (except the constant energy density $\Lambda$)  space-time described by the ({\it matter free}) 
EH-type NLSUSY GR action equipped with the positive potential minimum $V_{P.E.}=\Lambda >0$ 
(\ref{NLSUSYGR}) is unstable due to NLSUSY structure of tangent space and 
induce the phase transitions towards the true vacuum $V_{P.E.}=0$.  
It decays (called {\it Big Decay}) spontaneously to ordinary Riemann space-time with the NG fermion 
({\it superon} matter) described by the ordinary EH action with the cosmological term, the NLSUSY 
action for the $N$ NG fermions $superon$ and their gravitational interactions, 
which is called SGM action for superon-graviton(SG) model space-time.  
By expanding (\ref{NLSUSYGR}) around $e^a{}_\mu$  the SGM action is given as follows; 
\begin{equation}
L_{\rm SGM}(e,\psi) = - {c^4 \over {16 \pi G}} e \vert w_{\rm VA} \vert \{ R(e) + \Lambda + T(e, \psi)\} 
\label{SGM}
\end{equation}
where $R(e)$ is the Ricci scalar curvature of ordinary EH action, 
$T(e,\psi)$ represents highly nonlinear gravitational interaction terms of $\psi^i$, 
and $\vert w_{\rm VA} \vert = \det w^a{}_b = \det (\delta^a_b + t^a{}_b)$ 
is the NLSUSY invariant four dimensional volume \cite{VA}. 
We can easily see that 
the cosmological term in $L_{\rm NLSUSYGR}(w)$ of Eq.(\ref{NLSUSYGR}) 
(i.e. the constant energy density of ultimate space-time) 
mediated to the second term in SGM action (\ref{SGM}) reduces to the NLSUSY action \cite{VA}, 
$L_{\rm NLSUSY}(\psi) = -{1 \over {2 \kappa^2}} \vert w_{\rm VA} \vert$ 
in Riemann-flat $e_a{}^\mu(x) \rightarrow \delta_a^\mu$ space-time. 
Therefore, the arbitrary constant $\kappa$ of NLSUSY should be fixed to 
\begin{equation}
\kappa^{-2} = {{c^4 \Lambda} \over {8 \pi G}}. 
\label{kappa}
\end{equation}
The Big Decay is the phase transition ${sGL(4,R) \over GL(4,R)}$ of space-time. 
It  produces the  massless NG fermion, the ordinary Riemann space-time 
and a fundamental mass scale of the spontaneous breakdown of SUSY (SBS) 
depending on the $\Lambda$ and $G$ through the relation (\ref{kappa}).  
We will show that the effect of Big Decay is mediated to  the (low energy) particle physics 
in (asymptotic) Riemann-flat space-time. 
Note that  $L_{\rm SGM}(e,\psi)$ (\ref{SGM}) (massless superon-graviton model) 
preserves $V_{P.E.}=\Lambda>0$, which leaves the problem of identifying 
the true vacuum $V_{P.E.}=0$ for the massless fermion-graviton world.  

The nonlinear model sometimes can be related (converted) to the (equivalent) linear theory 
which is tractable.  
NLSUSY is also the case and the consequent linearized theory possesses the true vacuum $V_{P.E.}=0$, 
where the  (massless) particle spectrum is determined by the space-time sP symmetry and 
all particles and forces become composites of $\psi^i(x)$.  
We will investigate the low energy physics of NLSUSY GR through the NL/L SUSY relation. 
To see the (low energy) particle physics content in (asymptotic) Riemann-flat space-time 
we focus on $N=2$ SUSY in two dimensional space-time for simplicity, 
for in the SGM scenario $N=2$ case gives the minimal and realistic $N = 2$ LSUSY QED model \cite{STT}. 
By performing the systematic arguments  we can find the NL/L SUSY relation 
between the  $N=2$  NLSUSY model and $N=2$ LSUSY QED theory in Riemann-flat space-time \cite{ST3,ST4}; 
\begin{equation}
L_{N=2{\rm SGM}}(e,\psi) {\longrightarrow} 
L_{N=2{\rm NLSUSY}}(\psi) 
= L_{N=2{\rm LSUSYQED}}({\bf V},{\bf \Phi}) + [{\rm tot.\ der.\ terms}]. 
\label{NLSUSY-SUSYQED}
\end{equation}
In the relation (\ref{NLSUSY-SUSYQED}), the $N = 2$ NLSUSY action $L_{N=2{\rm NLSUSY}}(\psi)$ 
for the two (Majorana) NG-fermions superon $\psi^i$ $(i = 1, 2)$ is written in $d = 2$ as follows; 
\begin{eqnarray}
&\!\!\! &\!\!\! 
L_{N=2{\rm NLSUSY}}(\psi) 
\nonumber \\
&\!\!\! &\!\!\! 
\hspace*{5mm} = -{1 \over {2 \kappa^2}} \vert w_{\rm VA} \vert 
= - {1 \over {2 \kappa^2}} 
\left\{ 1 + t^a{}_a + {1 \over 2!}(t^a{}_a t^b{}_b - t^a{}_b t^b{}_a) 
\right\} 
\nonumber \\
&\!\!\! &\!\!\! 
\hspace*{5mm} = - {1 \over {2 \kappa^2}} 
\bigg\{ 1 - i \kappa^2 \bar\psi^i \!\!\not\!\partial \psi^i 
- {1 \over 2} \kappa^4 
( \bar\psi^i \!\!\not\!\partial \psi^i \bar\psi^j \!\!\not\!\partial \psi^j 
- \bar\psi^i \gamma^a \partial_b \psi^i \bar\psi^j \gamma^b \partial_a \psi^j ) 
\bigg\}, 
\label{NLSUSYaction}
\end{eqnarray}
where $\kappa$ is a constant with the dimension $({\rm mass})^{-1}$, 
which satisfies the relation (\ref{kappa}). 

On the other hand, in Eq.(\ref{NLSUSY-SUSYQED}), 
the $N = 2$ LSUSY QED action $L_{N=2{\rm SUSYQED}}({\bf V},{\bf \Phi})$ 
is constructed from a $N = 2$ minimal off-shell vector supermultiplet 
and a $N = 2$ off-shell scalar supermultiplet denoted ${\bf V}$ and ${\bf \Phi}$ respectively. 
Indeed, the most general $L_{N=2{\rm SUSYQED}}({\bf V},{\bf \Phi})$ in $d = 2$ 
with a Fayet-Iliopoulos (FI) $D$ term and Yukawa interactions, 
is given in the explicit component form as follows for the massless case; 
\begin{eqnarray}
L_{N=2{\rm SUSYQED}}({\bf V},{\bf \Phi}) 
&\!\!\! = &\!\!\!- {1 \over 4} (F_{ab})^2 
+ {i \over 2} \bar\lambda^i \!\!\not\!\partial \lambda^i 
+ {1 \over 2} (\partial_a A)^2 
+ {1 \over 2} (\partial_a \phi)^2 
+ {1 \over 2} D^2 
- {\xi \over \kappa} D 
\nonumber \\[.5mm]
& & 
+ {i \over 2} \bar\chi \!\!\not\!\partial \chi 
+ {1 \over 2} (\partial_a B^i)^2 
+ {i \over 2} \bar\nu \!\!\not\!\partial \nu 
+ {1 \over 2} (F^i)^2 
\nonumber \\[.5mm]
& & 
+ f ( A \bar\lambda^i \lambda^i + \epsilon^{ij} \phi \bar\lambda^i \gamma_5 \lambda^j 
- A^2 D + \phi^2 D + \epsilon^{ab} A \phi F_{ab} ) 
\nonumber \\[.5mm]
& & 
+ e \bigg\{ i v_a \bar\chi \gamma^a \nu 
- \epsilon^{ij} v^a B^i \partial_a B^j 
+ \bar\lambda^i \chi B^i 
+ \epsilon^{ij} \bar\lambda^i \nu B^j 
\nonumber \\[.5mm]
& & 
- {1 \over 2} D (B^i)^2 
+ {1 \over 2} A (\bar\chi \chi + \bar\nu \nu) 
- \phi \bar\chi \gamma_5 \nu \bigg\}
\nonumber \\[.5mm]
& & 
+ {1 \over 2} e^2 (v_a{}^2 - A^2 - \phi^2) (B^i)^2. 
\label{SQEDaction}
\end{eqnarray}
where $(v^a, \lambda^i, A, \phi, D)$ ($F_{ab} = \partial_a v_b - \partial_b v_a$) 
are the staffs of the {\it minimal} off-shell vector supermultiplet ${\bf V}$ 
representing $v^a$ for a $U(1)$ vector field, 
$\lambda^i$ for doublet (Majorana) fermions, 
$A$ for a scalar field in addition to $\phi$ for another scalar field 
and $D$ for an auxiliary scalar field, 
while ($\chi$, $B^i$, $\nu$, $F^i$) are the staffs of the (minimal) off-shell scalar supermultiplet ${\bf \Phi}$ 
representing $(\chi, \nu)$ for two (Majorana) fermions, 
$B^i$ for doublet scalar fields and $F^i$ for auxiliary scalar fields. 
Also $\xi$ in the FI $D$ term is an arbitrary dimensionless parameter turning to a magnitude of SUSY breaking mass, 
and $f$ and $e$ are Yukawa and gauge coupling constants with the dimension (mass)$^1$ (in $d = 2$), 
respectively. 
The $N = 2$ LSUSY QED action (\ref{SQEDaction}) can be rewritten in the familiar manifestly covariant form 
by using the superfield formulation (for further details see Ref.\cite{ST4}). 

In the relation (equivalence) of the two theories (\ref{NLSUSY-SUSYQED}), 
all component fields of $({\bf V},{\bf \Phi})$ of the $N = 2$ LSUSY QED action (\ref{SQEDaction}) 
are expressed uniquely as composites of the NG fermions $\psi^i$. 
We call them {\it {SUSY invariant relations}}, which terminate at ${\cal O}((\psi^{i})^4)$ in $d = 2$ and $N = 2$, 
\begin{equation}
({\bf V}, {\bf \Phi}) \sim (\xi, \xi^i) \kappa^{n-1} (\psi^j)^n \vert w_{\rm VA} \vert + \cdots \ (n = 0,1,2), 
\label{SUSYinv}
\end{equation}
where $\xi^i$ is arbitrary demensionless ($SO(2)$) overall parameters 
in the SUSY invariant relations for ${\bf \Phi}$ 
and $(\psi^j)^2 = \bar\psi^j \psi^j, \epsilon^{jk} \bar\psi^j \gamma_5 \psi^k, 
\epsilon^{jk} \bar\psi^j \gamma^a \psi^k$, etc. 
For example, some of SUSY invariant relations for ${\bf V}$  in the Wess-Zumino gauge are 
\[
\begin{array}{lll}
v^a &\!\!\! = - {i \over 2} \xi \kappa \epsilon^{ij} 
\bar\psi^i \gamma^a \psi^j \vert w \vert, &\!\!\! {}
\\[2mm]
\lambda^i &\!\!\! =  \xi \left[ \psi^i \vert w \vert 
- {i \over 2} \kappa^2 \partial_a 
\{ \gamma^a \psi^i \bar\psi^j \psi^j \vert w \vert \} \right]
\\[2mm]
A &\!\!\! =  {1 \over 2} \xi \kappa \bar\psi^i \psi^i \vert w \vert, &\!\!\ {}

\\
\cdots &\!\!\!  &\!\!\! {}

\end{array}
\]
which are promissing features for the SGM scenario. 
The explicit form \cite{ST3} of the SUSY invariant relations (\ref{SUSYinv}) are obtained {\it systematically} 
in the superfield formulation (for example, see Refs.\cite{ST4,IK,UZ}).  
The familiar LSUSY transformations among the component fields of the LSUSY supermultiplets $({\bf V},{\bf \Phi})$ 
are reproduced among the composite LSUSY supermultiplets by taking the NLSUSY transformations of the constituents $\psi^i$. 
We just mention that {\it four-NG fermion self-interaction terms (i.e. the condensation of $\psi^i$)} 
appearing only in the auxiliary fields $F^i$ of the scalar supermultiplet ${\bf \Phi}$ 
are crucial for the local $U(1)$ gauge symmetry of LSUSY theory in SGM scenario \cite{ST3,ST4}. 
The relation (\ref{NLSUSY-SUSYQED}) are shown explicitly (and systematically) 
by substituting Eq.(\ref{SUSYinv}) into the LSUSY QED action (\ref{SQEDaction}) \cite{ST3,ST4}. 

Now we briefly show the (physical) {\it true} vacuum structure of $N = 2$ LSUSY QED action (\ref{SQEDaction}) 
related (equivalent) to the $N = 2$ NLSUSY action (\ref{NLSUSYaction}) \cite{STL}. 
The vacuum is determined by the minimum of the potential $V_{P.E.}(A, \phi, B^i, D)$ in the action (\ref{SQEDaction}).    
The potential is  given by using the equation of motion for the auxiliary field $D$ as 
\begin{equation}
V_{P.E.}(A, \phi, B^i) = {1 \over 2} f^2 \left\{ A^2 - \phi^2 + {e \over 2f} (B^i)^2 
+ {\xi \over {f \kappa}} \right\}^2 + {1 \over 2} e^2 (A^2 + \phi^2) (B^i)^2 \ge 0, 
\label{potential}
\end{equation}
The configurations of fields corresponding to {\it true} vacua $V_{P.E.}(A, \phi, B^i)=0$  in $(A, \phi, B^i)$-space in the potential (\ref{potential}) 
are classified according to the signatures of the parameters $e, f, \xi, \kappa$. 

By adopting the simple parametrization $(\rho, \theta, \phi, \omega)$  for the vacuum configuration 
of $(A, \phi, B^i)$-space and by expanding the fields $(A, \phi, B^i)$ around the vacua, e.g. 
\[
\begin{array}{lll}
A &\!\!\! = - (k + \rho) \cos\theta \cos\varphi \cosh\omega, &\!\!\! {}
\\
\phi &\!\!\! = (k + \rho) \sinh\omega, &\!\!\! {}
\\
\tilde B^1 &\!\!\! = (k + \rho) \sin\theta \cosh\omega, &\!\!\ {}
\\
\tilde B^2 &\!\!\! = (k + \rho) \cos\theta \sin\varphi \cosh\omega. &\!\!\! {}
\\
A &\!\!\! =\phi=0  \ {\rm or} \  B^i = 0,  &\!\!\! {}
\label{paramet}
\end{array}
\]
we obtain the particle (mass) spectra of the linearized theory $N=2$ LSUSY QED.
We have found one of the vacua $V_{P.E.}(A, \phi, B^i)=0$ describes $N=2$ LSUSY QED containing \\[2mm]
\hspace*{5mm} 
one charged Dirac fermion ($\psi_D{}^c \sim \chi + i \nu$), \\
\hspace*{5mm} 
one neutral (Dirac) fermion ($\lambda_D{}^0 \sim \lambda^1 - i \lambda^2$), \\
\hspace*{5mm} 
one massless vector (a photon) ($v_a$), \\
\hspace*{5mm} 
one charged scalar ($\phi^c \sim \theta + i \varphi$), \\
\hspace*{5mm} 
one neutral complex scalar ($\phi^0 \sim \rho \ (+ i \omega)$), \\[2mm]
with masses $m_{\phi^0}^2 = m_{\lambda_D{}^0 }^2 = 4 f^2 k^2 = -{{4 \xi f} \over \kappa}$,  
$m_{\psi_D{}^c} ^2 = m_{\phi^c}^2 = e^2 k^2 =-{{\xi e^2} \over {\kappa f}}$, 
$m_{v_{a}} = 0$,
which are the composites of NG-fermions superon and the vacuum 
breaks SUSY alone  spontaneously (The local $U(1)$ is not broken. For further detailes, see \cite{STL,STLa}). 
For large $N$ case we can anticipate the large (broken) $SU(N)$ LSUSY with different large mass scales. 

Remarkably these arguments show that the true vacuum of ((asymptotic) Riemann-flat space-time of) $L_{N=2{\rm SGM}}(e,\psi)$ 
is achieved by the compositeness (eigenstates) of fields of the supermultiplets 
of {\it global} $N=2$ LSUSY QED. 
This phenomena may be regarded as the {\it relativistic} second order phase transition of massless superon-graviton system, 
which is dictated by the symmetry of space-time (analogous to the superconducting states achieved by the Cooper pair).
Here we should notice that R-parity (for $N \ge 2$) may not be a good quantum number in  the {\it true vacuum} of SGM scenario 
as seen from the particle spectra (without superpartners) mentioned above. 
These situations are very fabourable in constructing the consistent model with the recent LHC data  
which exclude the TeV scale SUSY breaking.  
As for the cosmological significances of $N = 2$ SUSY QED in the SGM scenario, 
the (physical) vacuum for the above model explains (predicts) simply the observed mysterious (numerical) relation 
between {\it the (dark) energy density of the universe} $\rho_D$ ($\sim {{c^4 \Lambda} \over {8 \pi G}}$) 
and {\it the neutrino mass} $m_\nu$, 
\begin{center}
\begin{equation}
\rho_D^{\rm obs} \sim (10^{-12} GeV)^4 \sim (m_\nu){}^4 
\sim {\Lambda \over G} \ (\sim {g_{\rm sv}}^2), 
\label{dark-m}
\end{equation}
\end{center}
provided $- \xi f \sim {\cal O}(1)$ and $\lambda_D{}^{0}$ is identified with  the neutrino, 
which gives a new insight into the origin of (small) mass \cite{STL,ST2} and 
produce the mass hielarchy by the factor ${e \over f}$ \big( $\sim {\cal O} \left({m_{e} \over m_{\nu}} \right)$ 
in case of $\psi_D{}^c$ as electron! \big). 

Furthermore, the neutral scalar field ${\phi^{0} (\sim  \rho) }$ with mass $\sim {\cal O}(m_\nu)$ 
of the radial mode in the vacuum configuration may be a candidate of {\it the dark matter}, 
for $N = 2$ LSUSY QED structure and the radial mode in the vacuum are preserved in the realistic large $N$ SUSY GUT model.  
(Note that $\omega$ in the model is a NG boson 
and disappears provided the corresponding local gauge symmetry is introduced as in the standard model.) 
These arguments show the potential of the SGM scenario which gives unified understandings for particle physics and cosmology. 
The no-go theorem for $N > 8$ SUSY may be overcome (irrelevant) in a sense that the linearized (equivalent) $N>8$ LSUSY theory would be 
{\it massive} theory with SSB. 

As for the magnitude of the bare coupling constant, by taking the more {\it general} auxiliary-field structure 
for the {\it general} off-shell vector supermultiplet (${v^a, \lambda^i, A, \phi, D,}$
${M^{ij}, \Lambda, C}$) \cite{ST5} and establishing the NL/L SUSY 
relation we have shown that 
{\it the magnitude of the bare (dimensionless) gauge coupling constant $e$} 
(i.e. the fine structure constant $\alpha = {e^2 \over 4\pi}$) is expressed (determined) 
in terms of {\it vacuum values of auxiliary-fields} \cite{ST5}: 
\begin{equation}
e = {\ln({\xi^i{}^2 \over {\xi^2 - 1}}) \over 4 \xi_C}, 
\label{f-xi}
\end{equation}
where $e$ is the bare gauge coupling constant, $\xi$, $\xi^i$ and $\xi_C$ are 
the vacuum-value scale parameters (the relative magnitudes in the NL/L SUSY relation)
of auxiliary-fields of the {\it general} off-shell supermultiplet in $d=2$.  
This mechanism is natural and very favourable for SGM scenario as a {\it theory of everything}. 

\section{Conclusions}
We have proposed a new paradigm for describing the unity of nature, 
where the ultimate shape of nature is new unstable $(V_{P.E}>0)$ space-time 
described by the NLSUSY GR action $L_{\rm NLSUSYGR}(w)$ in the form of 
the free EH action for empty space-time with the constant energy density. 
Big Decay of new space-time $L_{\rm NLSUSYGR}(w)$ creates 
ordinary Riemann space-time with {\it massless} spin-${1 \over 2}$ superon 
described by the SGM action $L_{\rm SGM}(e,\psi)$ with $(V_{P.E}>0)$ 
and ignites Big Bang of space-time and matter accompanying the dark energy (cosmological constant). 
Interestingly on Riemann-flat tangent space (in the local frame), 
the familiar renormalizable LSUSY theory emerges on the true vacuum $(V_{P.E}=0)$ of 
SGM action  $L_{\rm SGM}(e,\psi)$ as mssless composite-eigenstates of superon. 
We can anticipate in the true vacuum the larger gauge symmetries and 
the consequent different mass scales  for the NL/L SUSY relation for the larger $N$. 
We have seen that the physics before/of the Big Bang may play crucial roles 
for understanding unsolved problems of the universe and the particle physics.   \par  
In fact, we have shown explicitly that $N=2$ LSUSY QED theory as the realistic $U(1)$ gauge theory  
emerges in the physical field configurations 
on the true vacuum of $N=2$ NLSUSY theory on Minkowski tangent space-time, which gives new insights into 
the origin of mass and the cosmological problems.  
The cosmological implications of the composite SGM scenario seems promissing but deserve further studies.   \par
Remarkably the physical particle states of $N=2$ LSUSY as a whole 
look  the similar structure to the lepton sector of ordinary SM with the local $U(1)$ 
and the implicit global $SU(2)$ \cite{STT} disregarding the R-parity, i.e. 
without the trivial superpartner. 
Such SUSY breaking mechanism may allow the SUSY model construction without introucing  ${a priori}$ the superpartners, 
which is compatible with the recent LHC data excluding the low (TeV) mass superpartners. 
(Note that the scalar mode  $\omega$ is a NG boson and disappears provided the corresponding local gauge symmetry 
is introduced.) 
We anticipate that the physical consequences obtained in $d=2$ hold in $d=4$ as well, 
for the both have the similar structures of on-shell helicity states of $N=2$ supermultiplet 
though scalar fields and off-shell (auxiliary field) structures are modified (extended). 
However, the similar investigations in $d = 4$ are urgent for the realistic model building based upon SUSY. \par  
The extension to large $N$, especially to $N = 5$ 
is important for {\it superon\ quintet\ hypothesis} of SGM scenario 
with ${N = \underline{10} = \underline{5}+\underline{5^{*}}}$ for equipping the $SU(5)$ GUT structure \cite{KS}  
and to  $N = 4$ may shed new light on the mahematical structures of 
the anomaly free non-trivial $d=4$ field theory. 
($N=10$ SGM predicts double-charge heavy lepton state $E^{2+}$ and new one neutral singlet massive vector state \cite{KS0}). 
Further investigations on the spontaneous symmetry breaking for $N \geq 2$ SUSY remains to be studied.  
It may be helpful to point out the similarity between (high $T_{c}$) superconductivity or superfluidity  and SGM scenario. 
The structure (symmetry)   of the bulk (space-time)   determines the resulting spectra.  \par
Linearizing SGM action ${L_{\rm SGM}(e,\psi)}$ on curved space-time, 
which elucidates the topological structure of space-time \cite{SK}, is a challenge. 
The corresponding NL/L SUSY relation will give the supergravity (SUGRA) \cite{FNF,DZ} 
analogue with the vacuum which breaks SUSY spontaneously. \par  

Locally homomorphic non-compact groups $SO(1,3)$ and $SL(2,C)$ for  
space-time degrees of freedom are analogues of compact groups  $SO(3)$ and $SU(2)$ 
for gauge degrees of freedom of 't Hooft-Polyakov monopole. 
They are special, because they are unique homomorphic groups among 
$SO(1,D-1)$ and $SL(d,C)$, i.e. 
\begin{equation}
{D(D-1) \over 2}=2(d^{2}-1) 
\label{homoml}
\end{equation}
holds for only 
\begin{equation}
D=4,\ {} d=2. 
\label{so-sl}
\end{equation}
NLSUSYGR/SGM scenario predicts {\it four} dimensional space-time. ($D=d=1$ is trivial.)

Finally we just mention that NLSUSY GR and the subsequent SGM scenario 
for the spin-${3 \over 2}$ NG fermion \cite{ST1,Baak} is in the same scope. \par

Our discussion shows that considering the physics before/of the Big Bang may be significant for 
cosmology and the (low energy) particle physics as well, where SUSY and its spontaneous breakdown 
may play crucial roles and leave evidences which can be tested in the low energy.
\vspace{10mm}

The authors would like to thank Professor T. Okano, Department of Mathematics of SIT 
for giving the proof of Eq. (13).

\vspace{1.5cm}

\newpage

%
\newcommand{\NP}[1]{{\it Nucl.\ Phys.\ }{\bf #1}}
\newcommand{\PL}[1]{{\it Phys.\ Lett.\ }{\bf #1}}
\newcommand{\CMP}[1]{{\it Commun.\ Math.\ Phys.\ }{\bf #1}}
\newcommand{\MPL}[1]{{\it Mod.\ Phys.\ Lett.\ }{\bf #1}}
\newcommand{\IJMP}[1]{{\it Int.\ J. Mod.\ Phys.\ }{\bf #1}}
\newcommand{\PR}[1]{{\it Phys.\ Rev.\ }{\bf #1}}
\newcommand{\PRL}[1]{{\it Phys.\ Rev.\ Lett.\ }{\bf #1}}
\newcommand{\PTP}[1]{{\it Prog.\ Theor.\ Phys.\ }{\bf #1}}
\newcommand{\PTPS}[1]{{\it Prog.\ Theor.\ Phys.\ Suppl.\ }{\bf #1}}
\newcommand{\AP}[1]{{\it Ann.\ Phys.\ }{\bf #1}}

\end{document}